\documentclass[twocolumn,showpacs,preprintnumbers,amsmath,amssymb, superscriptaddress, prb]{revtex4}

\usepackage{graphicx}
\usepackage{dcolumn}
\usepackage{bm}

\usepackage{color}
\definecolor{red}{rgb}{1,0,0}

\definecolor{blue}{rgb}{0,0,1}

\definecolor{green}{rgb}{0,1,0}

\begin{document}
\preprint{APS}

\title{Magnetic phase diagram of the multiferroic FeTe$_2$O$_5$Br}
\author{M. Pregelj}
\email{matej.pregelj@ijs.si}
\affiliation{Institute "Jozef Stefan", Jamova 39, 1000 Ljubljana, Slovenia}
\author{A. Zorko}
\affiliation{Institute "Jozef Stefan", Jamova 39, 1000 Ljubljana, Slovenia}
\author{O. Zaharko }
\affiliation{Laboratory for Neutron Scattering, PSI, CH-5232 Villigen, Switzerland}
\author{Z. Kutnjak}
\affiliation{Institute "Jozef Stefan", Jamova 39, 1000 Ljubljana, Slovenia}
\author{M. Jagodi\v{c}}
\affiliation{Institute of Mathematics, Physics and Mechanics, Jadranska 19, 1000 Ljubljana, Slovenia}
\affiliation{EN$\rightarrow$FIST Centre of excellence, Dunajska 156, 1000 Ljubljana, Slovenia}
\author{Z. Jagli\v{c}i\' {c}}
\affiliation{Institute of Mathematics, Physics and Mechanics, Jadranska 19, 1000 Ljubljana, Slovenia}
\affiliation{Faculty of Civil and Geodetic Engineering, University of Ljubljana, Jamova 2, 1000 Ljubljana, Slovenia}
\author{H. Berger}
\affiliation{Institute of Physics of Complex Matter, EPFL, 1015 Lausanne, Switzerland}
\author{M. de Souza}
\affiliation{Physikalisches Institut, Goethe-Universit\"at Frankfurt(M), SFB/TR49, D-60438 Frankfurt am Main, Germany}
\author{C. Balz}
\affiliation{Physikalisches Institut, Goethe-Universit\"at Frankfurt(M), SFB/TR49, D-60438 Frankfurt am Main, Germany}
\author{M. Lang}
\affiliation{Physikalisches Institut, Goethe-Universit\"at Frankfurt(M), SFB/TR49, D-60438 Frankfurt am Main, Germany}
\author{D. Ar\v{c}on}
\affiliation{Institute "Jozef Stefan", Jamova 39, 1000 Ljubljana, Slovenia}
\affiliation{Faculty of Mathematics and Physics, University of Ljubljana, Jadranska 19, 1000 Ljubljana, Slovenia}
\date{\today}

\begin{abstract}
The low-temperature magnetic phase diagram of the multiferroic system FeTe$_2$O$_5$Br  down to 300\,mK and up to 9\,T is presented. Short-range magnetic correlations within the crystal layers start to develop already at $\sim$50\,K, i.e., far above $T_{N1} \sim$ 11.0\,K, where the system undergoes a magnetic phase transition into the high-temperature incommensurate (HT-ICM) phase. Only 0.5\,K lower, at $T_{N2}$, the system undergoes a second phase transition into the low-temperature incommensurate amplitude-modulated (LT-ICM) phase accompanied by a spontaneous electric polarization. When the magnetic field is applied, the transition temperatures shift depending on the field orientation. In the case of $B||b$ and $B >$ 4.5\,T, the HT-ICM phase disappears along with the electric polarization in the LT-ICM phase. The field dependence of the magnetic transition temperatures is explained in the context of the magnetic susceptibility behavior. Similarities and differences between the novel amplitude-modulated and well-established helicoidal magnetoelectrics are discussed.
\end{abstract}

\maketitle

\section{Introduction}
A magnetically driven ferroelectric response \cite{ScottNature, Kimura, Nmat07, Lawes05, Kenzelmann05} has been almost exclusively observed in incommensurate (ICM) states with broken inversion symmetry, where non-centrosymmetric lattice distortions and ferroelectric order are induced through exchange-striction, \cite{ES,ES1,ES2,ES3} inverse Dzyaloshinskii-Moriya \cite{IDM} or spin current \cite{SC} mechanisms. Since complex ICM magnetic orderings without inversion symmetry are often provoked by magnetic frustration, resulting from competing exchange interactions on a lattice of localized spins, low-dimensional systems with triangular geometries are considered as prominent candidates for novel magnetoelectric materials.

One of the synthesis strategies that has proved to be very successful in the search of such compounds, is to use lone-pair cations and mix them with a transition metal in the presence of halogen ions.\cite{Johnsson} This way it is very likely that the number of the superexchange pathways between the magnetic ions is reduced and a geometrically frustrated low-dimensional structure is formed. On the other hand, lone-pair electrons were also recognized as carriers of the electric polarization in numerous ferroelectric materials, such as for instance Bi$^{3+}$ in BiMnO$_3$. \cite{Seshadri} In fact, it is generally accepted that since lone-pair electrons are stereochemically active, they can be easily polarized. They are thus considered as a primary driving force behind the off-center structural distortions essential for the formation of electric polarization in magnetic materials. For these reasons, they seem to be convenient candidates to induce both, magnetic frustration as well as electric polarization, and may consequently lead to a strong coupling between magnetic and electric orders.

This assumption has been proven correct by the discovery of the magnetoelectric coupling in FeTe$_2$O$_5$Br,\cite{Pregelj} which is an exemplary product of the above research directives. This system has a crystal structure that implies both magnetic frustration and reduced dimensionality. It adopts a layered structure, where individual layers consist of geometrically frustrated iron tetramer units [Fe$_4$O$_{16}$]$^{20-}$ coupled through the [Te$_4$O$_{10}$Br$_2$]$^{6-}$ groups.\cite{Becker} The negative Curie-Weiss temperature $\theta_{CW}$ = $-$98\,K, determined from susceptibility measurements,\cite{Becker} implies strong antiferromagnetic (AFM) interactions between the Fe$^{3+}$ ($S$ = 5/2) moments. The strongly suppressed N\'{e}el temperature, $T_N \sim$ 10\,K, suggests that the exchange interactions are frustrated.\cite{Becker} Our recent investigation \cite{Pregelj} revealed that at $T_N$ = 10.6(1)\,K the system undergoes a transition into an ICM amplitude-modulated magnetic structure [with magnetic wave vector $k$ = (1/2 0.463 0)], which is accompanied by a spontaneous electric polarization, pointing perpendicular to $k$ and to the magnetic moments. The ferroelectricity was ascribed to the polarization of the Te$^{4+}$ lone-pair electrons, while the magnetoelectric effect was argued to be due to sliding of neighboring amplitude-modulated waves, which induces the exchange-striction of the Fe-O-Te-O-Fe intercluster exchange bridges.

We stress that although the novel ICM amplitude-modulated structure in FeTe$_2$O$_5$Br differs from helical and cycloidal magnetic orders typically found in other multiferroics with spin-order induced ferroelectricity, the coupling mechanism is still described with two complex magnetic order parameters. One of the fundamental questions is whether this similarity leads also to a similar phase diagram, i.e., how does the applied magnetic field affect the ferroelectric properties, and vice versa, how the electric field affects the magnetic properties of the system with ICM amplitude-modulated magnetic structure. A small value of the electric polarization, on one hand, and sizable Fe$^{3+}$ magnetic moments on the other, imply that the magnetoelectric effect should be more easily induced by applying the magnetic field than the electric field. We therefore explored the magnetic phase diagram and the influence of the applied magnetic field on the magnetic and electric properties of the FeTe$_2$O$_5$Br system. In particular, we were interested in seeing whether the external magnetic field can suppress or even switch the electric polarization. Our detailed study includes a variety of complementary experimental techniques; i.e., specific heat, magnetic susceptibility, neutron diffraction, dielectric and thermal expansion measurements were performed down to 300\,mK and up to 9\,T.

In this paper we show that short-range magnetic correlations within the crystal layers persist up to $\sim$50\,K, while long-range magnetic ordering sets in at $T_{N1}$ = 11.0(1)\,K, when the HT-ICM phase is established. At $T_{N2}$ = 10.5(1)\,K a second transition occurs and LT-ICM phase accompanied by a spontaneous electric polarization emerges. In an external magnetic field the transition temperatures strongly depend on the field strength as well as its orientation. In the case of $B||b$ and $B >$ 4.5\,T, the HT-ICM phase disappears along with the electric polarization in the LT-ICM phase.

\section{Experimental}
High quality single crystals of FeTe$_2$O$_5$Br were grown by the standard chemical vapor phase method, reported elsewhere.\cite{Becker}

Specific heat measurements were performed in the temperature range between 20\,K and 2\,K and applied magnetic fields between 0 to 9\,T. Zero-field measurements in the temperature range between 15\,K and 0.3\,K were performed using a closed-cycle He-3 cryostat. All measurements were performed on the commercial Quantum Design PPMS setup.

Magnetic susceptibility ($\chi = M/H$) measurements between 300\,K and 2\,K in the applied magnetic field up to 5\,T were performed with Quantum Design MPMS XL-5 SQUID magnetometer using a closed-cycle cryostat.

Temperature dependence of the complex dielectric constant, $\epsilon^*(T, B) = \epsilon'(T, B)-i\epsilon''(T, B)$ was measured as a function of temperature and frequency $\nu$ by using an HP4282A precision LCR meter. The dielectric constant was scanned at few frequencies between 20\,Hz and 1\,MHz on cooling or heating the sample with the typical cooling/heating rates of 10\,K/h in the various dc bias electric fields ranging from 0-3\,kV/cm. The excitation electric ac field of 100-400\,V/cm was applied along the $a^*$-, $b$- and $c$-axes. The quasistatic polarization $P$ was determined by electrometer charge-accumulation measurements as described in Refs. \onlinecite{eps1} and \onlinecite{eps2} in a field-cooling run. Here the bias field of 10\,kV/cm was used, which was several times higher than the coercive field ($\sim$1\,kV/cm) in order to obtain saturated spontaneous polarization. Zero-field ac dielectric measurements and ac dielectric measurements in the dc electric bias field were performed in an Oxford continuous-flow liquid-helium cryostat. The ac dielectric measurements in the dc magnetic fields up to 5\,T were performed using the MPMS's cryostat as well as the MPMS's temperature and magnetic field control.

Measurements of the magnetic reflections between 50\,K and 1.5\,K under the applied magnetic field up to 6\,T were performed on a 5$\times$4$\times$1\,mm$^3$ single crystal using the single crystal diffractometer TriCS ($\lambda$ = 2.32\,\AA), upgraded with an Oxford superconducting magnet at the Swiss Neutron Spallation Source, Paul Scherrer Institute, Switzerland.

High-resolution thermal expansion measurements were performed using a capacitive dilatometer\cite{Pott} capable of resolving length changes $\Delta l \geq 0.01$\,\AA. The data were taken during  warming up, by employing a sweep rate of 1.5\,K/h. The thermal expansion data were corrected for the thermal expansion of the dilatometer cell.

\section{Results}

\subsection{Phase transitions for $B||a^*$}

\begin{figure}[!]
\includegraphics[width=86mm,keepaspectratio=true]{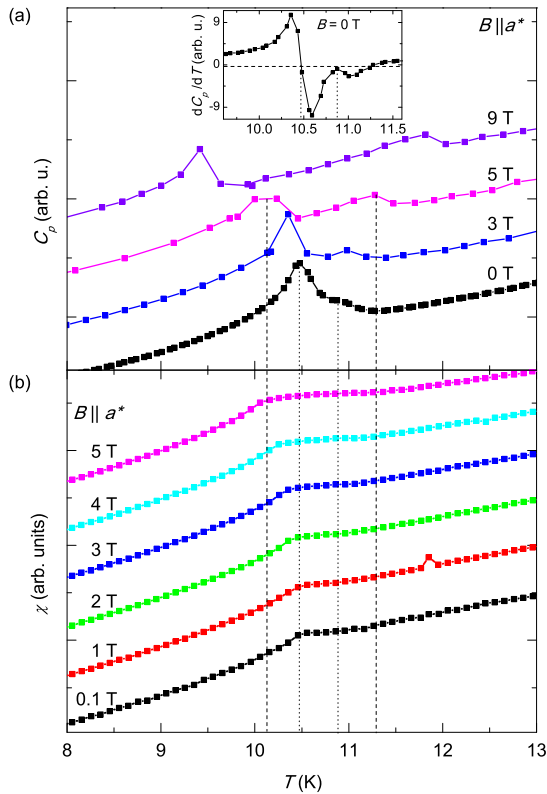}
\caption {(a) Temperature dependence of the specific heat and (b) magnetic susceptibility in magnetic fields applied along the $a^*$-axis. Dashed and doted lines indicate the position of $T_{N1}$ and $T_{N2}$ at 0\,T and 5\,T, respectively. Data measured in different fields have been shifted vertically for clarity.}
\label{fig1}
\end{figure}

Although initial zero-field experiments on FeTe$_2$O$_5$Br indicated a single magnetic transition at $T_N$ = 10.6\,K, \cite{Becker, Pregelj} specific heat, $C_p$, measurements for $B||a^*$ (Fig.\,\ref{fig1}a) reveal that on cooling the FeTe$_2$O$_5$Br system undergoes two consecutive transitions at $T_{N1}(B=0)=11.0(1)$\,K and $T_{N2}(B=0)=10.5(1)$\,K. The two transitions are indicated by two anomalies. At $B \geq$ 3\,T they are well separated and their shapes imply that the upper transition has a step-like nature, while the lower one is a broadened $\lambda$-like transition. Two consecutive transitions were similarly observed in the FeTe$_2$O$_5$Cl system, \cite{Becker} where transitions are $\sim$1.5\,K apart already in zero-field. In FeTe$_2$O$_5$Br, however, the transitions at 0\,T are separated by only 0.5\,K and can be distinguished only from the d$C_p$/d$T$ plot (inset of Fig.\,\ref{fig1}a), which is why they were overlooked in previous studies. \cite{Becker, Pregelj} With increasing magnetic field the upper anomaly, at $T_{N1}$, moves to higher, while the lower one, at $T_{N2}$, shifts to lower temperatures. The shifts are quite significant, i.e., at 9\,T $T_{N1}$ = 11.8(1)\,K and $T_{N2}$ = 9.4(1)\,K.

To address the magnetic character of the observed phases we performed temperature scans of the magnetic susceptibility, $\chi(T)$, at fixed magnetic fields of 0.1, 1, 2, 3, 4, and 5\,T (Fig.\,\ref{fig1}b). Only the anomaly at $T_{N2}$ is observed in $\chi(T)$. With increasing field this anomaly shifts to lower temperatures, reflecting the behavior of $T_{N2}$ and thereby suggesting that magnetic long-range order may not exists above $T_{N2}$.

\begin{figure}[!]
\includegraphics[width=86mm,keepaspectratio=true]{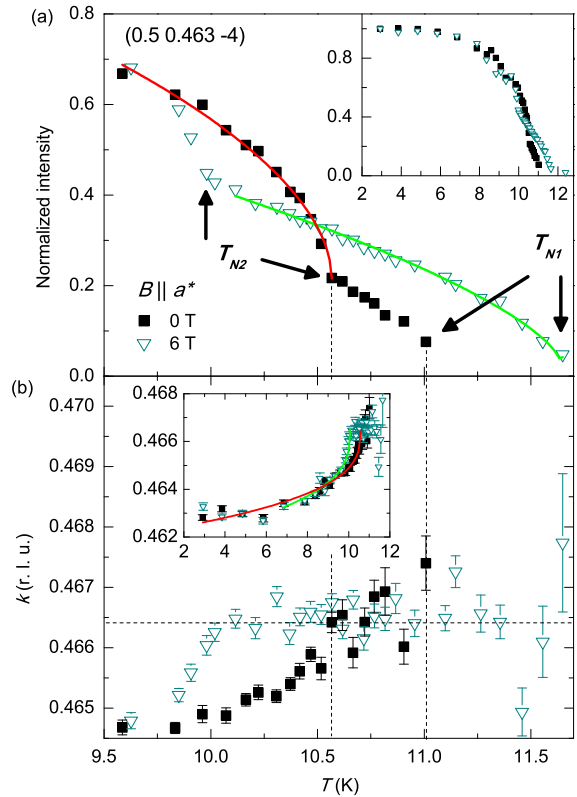}
\caption {(a) Temperature dependence of the (0.5 0.463 -4) magnetic peak intensity measured at the magnetic fields 0\,T and 6\,T applied along the $a^*$-axis. Inset:  temperature dependence of the peak intensity on an expanded temperature range. (b) Temperature dependence of the peak position with respect to the $k$ direction in reciprocal space. Inset: temperature dependence below $T_{N1}$ on an expanded temperature range. Solid lines in (a) and inset to (b) represent fits to the $(T_N - T)^\beta$ power law.}
\label{fig2}
\end{figure}

To clarify the nature of the phase between $T_{N1}$ and $T_{N2}$, a series of single-crystal neutron diffraction experiments were performed. The detailed temperature-dependence scans of several magnetic peaks at $B$ = 0\,T reveal that they emerge already at  $T_{N1}$ = 11.0(1)\,K,\cite{Zaharko} while an anomaly in their intensities is found at  $T_{N2}$ =  10.5(1)\,K [see Fig.\,\ref{fig2}a - for clarity only the behavior of the (0.5 0.463 -4) magnetic peak intensity is shown]. This proves that long-range magnetic ordering indeed exists in both low-temperature phases. On cooling, the system first undergoes a transition from paramagnetic to a HT magnetic phase at $T_{N1}$, signified by the emergence of the magnetic reflections. Approximately 0.5\,K lower (at $T_{N2}$), a LT magnetic phase, indicated by the inclination in the magnetic peak intensities, is stabilized. When a magnetic field is applied, the temperature interval between $T_{N1}$ and $T_{N2}$ increases exactly as anticipated from the specific heat measurements (Fig.\,\ref{fig1}a). The behavior of the magnetic peak intensities in the HT and the LT phases is obviously intrinsically different. The increase of the intensity with decreasing temperature in the HT phase is surprisingly slow. The difference is even more pronounced when looking at the magnetic peak position (Fig.\,\ref{fig2}b), which appears to be temperature independent [locked to $k$ = 0.4665(3)] in the HT phase, while below $T_{N2}$ it gradually shifts to lower $k$ values. For instance, the (0.5 0.463 -4) magnetic peak shifts for 0.004(1)\,r.l.u. (inset of Fig.\,\ref{fig2}b). As $k$ is incommensurate in both phases, we name the phases as the HT incommensurate (HT-ICM) and the LT incommensurate (LT-ICM) magnetic phases. The intensities of the collected magnetic peaks in the HT-ICM phase, however, are too weak for a successful refinement, as expected for a very small ordered component of the magnetic moment.

\begin{figure}[!]
\includegraphics[width=86mm,keepaspectratio=true]{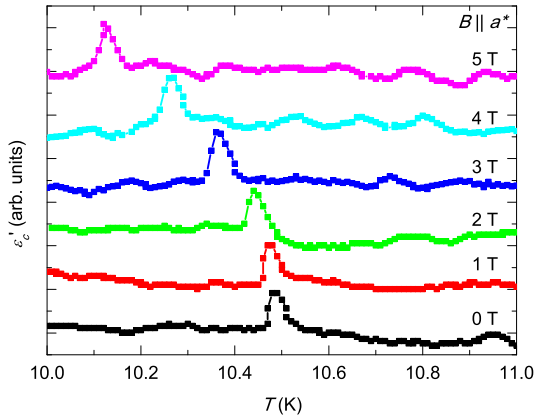}
\caption {Temperature dependence of the dielectric constant, $\epsilon$, measured for $E||c$ and magnetic fields along the the $a^*$-axis ranging between 0\,T and 5\,T. Data have been shifted vertically for clarity.}
\label{fig2a}
\end{figure}

In order to investigate the dielectric nature of the two phases as well as the electric response of the system to the magnetic field $B || a^*$, the dielectric constant, $\epsilon$, was measured (Fig.\,\ref{fig2a}). Since the electric polarization was found to be the largest along the $c$-axis, \cite{Pregelj} the measurements were performed in the corresponding electric field orientation ($E || c$). In contrast to the specific heat measurements, $\epsilon$ exhibits only one anomaly, which in the applied magnetic field precisely follows the behavior of $T_{N2}$. This clearly reveals that only the LT-ICM phase is ferroelectric. Moreover, it implies that the symmetry of the HT-ICM magnetic ordering probably prohibits ferroelectricity, and hence differs from the LT-ICM amplitude-modulated order explored in Ref.  \onlinecite{Pregelj}.

\subsection{Phase transitions for $B||c$}

\begin{figure}[!]
\includegraphics[width=86mm,keepaspectratio=true]{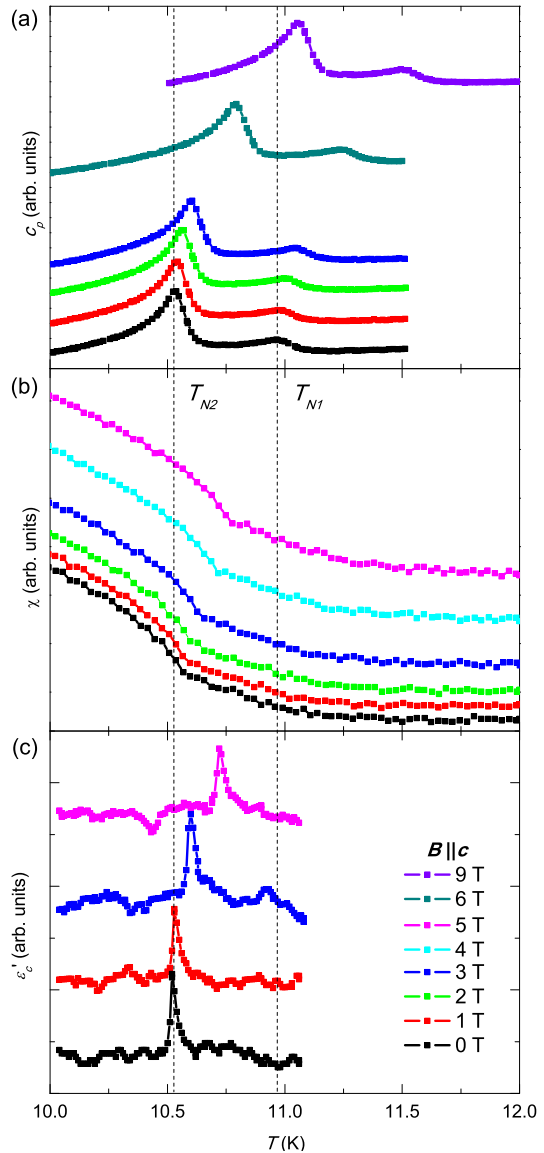}
\caption {(a) Temperature dependence  of the specific heat measured at several different magnetic fields applied along the $c$-axis. (b) Temperature dependence  of the magnetic susceptibility for magnetic fields applied along the $c$-axis. (c) Temperature dependence of the dielectric constant ($E||c$) measured in magnetic fields applied along the $c$-axis. Data have been shifted vertically for clarity.}
\label{fig7}
\end{figure}

Intrigued by the impact of $B||a^*$ on the phase transition temperatures, we continued our investigation by exploring the influence of a magnetic field applied along the $c$-axis. We first measured the temperature dependence of the specific heat in applied magnetic fields (Fig.\,\ref{fig7}a). For this orientation of the sample, both transitions are very well resolved already in zero-field. With increasing magnetic field, both transitions simultaneously shift to higher temperatures, i.e. $T_{N2}$ and $T_{N1}$ shift from 10.5(1)\,K and 11.0(1)\,K at 0\,T, to 11.1(1)\,K and 11.5(1)\,K at 9\,T, respectively, keeping the width of the HT-ICM temperature interval virtually unchanged.

To confirm these results, we performed magnetic susceptibility measurements (Fig.\,\ref{fig7}b). In contrast to the $C_p$ measurements, again only $T_{N2}$ can be clearly observed as a sharp increase of $\chi$, while the anomaly at $T_{N1}$ is less pronounced. We note that both magnetic phases appreciate the magnetic field applied along the $c$-axis, reflected in the increase of the magnetic transition temperatures.

Finally, we measured the dielectric response in $E||c$ for $B||c$. In agreement with the observed behavior of $C_p$ and $\chi$, the increasing magnetic field shifts the peak in $\epsilon_c$ to higher temperatures, implying that this orientation of the magnetic field stimulates ferroelectric ordering. Additionally, we stress that the height of the dielectric peak is not affected, suggesting that the magnitude and the orientation of the electric polarization are preserved.

\subsection{Phase transitions for $B||b$}

\begin{figure}[!]
\includegraphics[width=86mm,keepaspectratio=true]{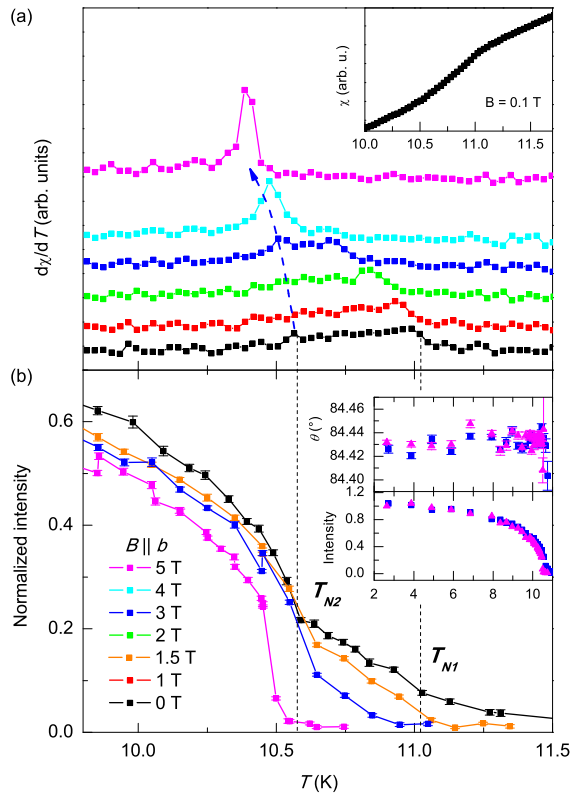}
\caption {(a) Derivative of the temperature dependence  of the magnetic susceptibility, d$\chi$/d$T$, indicating two successive magnetic transitions. Inset: $\chi(T)$ measured at 0.1\,T. (b) Temperature dependence of the (0.5 0.463 -4) magnetic peak intensity for a magnetic field along the $b$-axis. Inset: evolution of the position (up) and intensity (down) of the peak at 3\,T and 5\,T.}
\label{fig4}
\end{figure}

Last we measured the response of the system to $B||b$. Due to a specific plate-like shape of the crystals the specific heat measurements were not possible for this orientation. Magnetic transitions could not be clearly distinguished from the magnetic susceptibility (inset to Fig.\,\ref{fig4}a). We therefore show in Fig.\,\ref{fig4}a the derivatives of the temperature dependence of the magnetic susceptibility measured in different fields. Apparently, $T_{N1}$ lowers with increasing magnetic field, while the decrease of $T_{N2}$ is significantly less pronounced. Eventually, at $\sim$5\,T, both transitions seem to overlap, suggesting that the HT-ICM phase might have disappeared.

To obtain complementary information about the impact of $B||b$ on the magnetic properties of the system, we performed neutron diffraction experiments also for this crystal orientation. Detailed measurements in fields of 1.5, 3 and 5\,T (Fig.\,\ref{fig4}b) nicely corroborate the magnetic susceptibility results. Actually, here the extinction of the HT-ICM phase is even more evident. The HT-ICM temperature interval, indicated by the extraordinary, almost linear, temperature dependence of the magnetic peak intensity, is reduced with increasing field up to 5\,T, where no sign of the HT-ICM phase is left. On the other hand, the temperature dependence  of the magnetic peak intensity and position below $T_{N2}$ (insets to Fig.\,\ref{fig4}b) do not show any noticeable change between 3\,T and 5\,T, implying that the LT-ICM phase has not changed.

\begin{figure}[!]
\includegraphics[width=86mm,keepaspectratio=true]{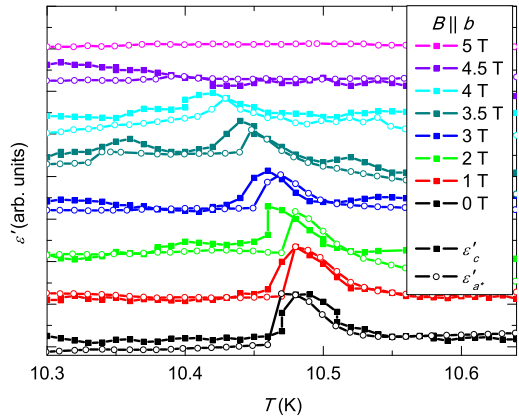}
\caption {Temperature dependence of the dielectric constant $\epsilon'$ for $E||c$ (solid squares) and $E||a^*$ (empty circles) measured  in magnetic fields applied along the $b$-axis. Data have been shifted vertically for clarity.}
\label{fig3}
\end{figure}

The intriguing question is then what happens with the electric polarization. In Fig.\,\ref{fig3} we show the temperature dependence  of the dielectric constant for $E||c$, measured in different magnetic fields $B||b$ up to 5\,T. The peak in the dielectric constant only marginally shifts up to $\sim$3\,T (similarly to $T_{N2}$ determined from $\chi$). However, with further increasing magnetic field, the peak starts to collapse. It becomes very weak at 4\,T and completely disappears at 4.5\,T. This can be either due to the suppression of the $c$ component of the electric polarization, $P_c$, or alternatively due to the rotation of the spontaneous polarization away from the $c$-axis. In order to distinguish between the two possibilities, we measured the dielectric constant also for $E||a^*$, while keeping $B$ along $b$. Evidently, the small $a^*$ component of the electric polarization \cite{Pregelj} results in a similar anomaly as found for $E||c$, which again disappears at $\sim$4.5\,T (Fig.\,\ref{fig3}). For $E||b$ no anomaly in $\epsilon$ has been found in the entire temperature range up to 5\,T (not shown here). This strongly suggests that the macroscopic electric polarization actually disappears for $B >$ 4.5\,T. Obviously, $B||b$ does not influence the electric polarization and the long-range magnetic ordering in the same way, as at 5\,T the first gets suppressed, while the second does not (Fig.\,\ref{fig3}).

\subsection{Lattice distortion - thermal expansion measurements}

A particularly sensitive probe for studying phase transitions is provided by measurements of the uniaxial thermal expansion coefficient, $\alpha_i(\textit{T})=\textit{l}^{-1}(\partial \textit{l}/\partial \textit{T})$, where $i$ indicates the uniaxial direction. In fact, lattice effects are naturally expected and observed  at a ferroelectric transition (see, e.g.\,, Ref.\,\onlinecite{Mariano08}), as atomic displacements, breaking the inversion symmetry, are prerequisite for ferroelectricity to occur.

\begin{figure}[!ht]
\begin{center}
\includegraphics[width=86mm,keepaspectratio=true]{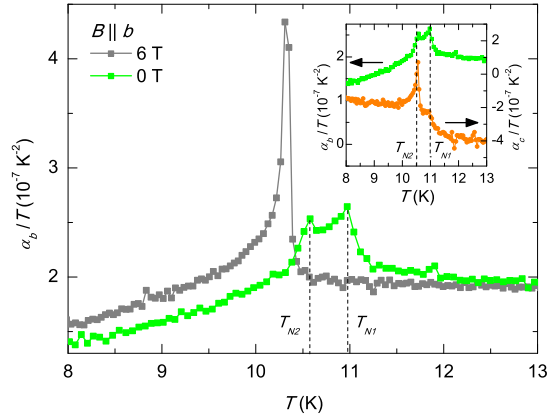}
\end{center}
\caption{Main panel: Linear thermal expansion coefficient $\alpha_b$
for FeTe$_2$O$_5$Br as $\alpha_b/T$ \emph{vs.} $T$. Data were taken in zero field and at a magnetic field of 6\,T for $B||b$. $T_{N1}$  and $T_{N2}$ indicate the transition temperatures of the magnetic and ferroelectric transitions at 11.0(1)\,K and 10.6(1)\,K, respectively, as discussed in the main text. Inset: $\alpha_b/T$ \emph{vs.} $T$ (left scale) together with $\alpha_c/T$ \emph{vs.} $T$ data (right scale).}\label{thermalexpansion-b}
\end{figure}

\begin{figure}[!ht]
\begin{center}
\includegraphics[width=86mm,keepaspectratio=true]{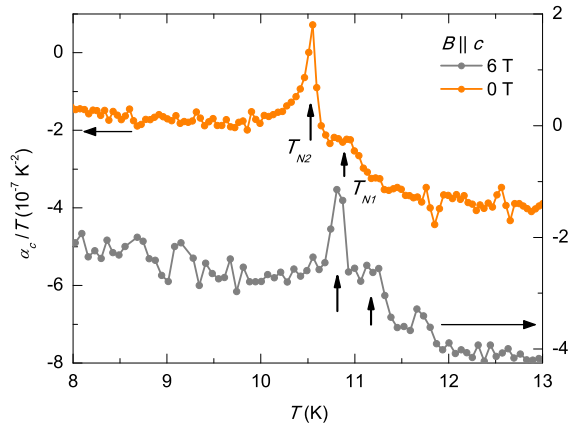}
\end{center}
\caption{Linear thermal expansion coefficient along the $c$-axis, $\alpha_c$, as $\alpha_c/T$ $vs.$ $T$ at zero-field (left scale) and at a magnetic field of 6\,T for $B||c$ (right scale).}\label{thermalexpansion-c}
\end{figure}

In Figs.\,\ref{thermalexpansion-b} and \ref{thermalexpansion-c}, we show the results of the uniaxial thermal expansion coefficient along the $i$ = \emph{b}- and \emph{c}-axes, respectively, in $\alpha_i$/$T$ \emph{vs.}\,$T$ plots. In zero field, $\alpha_b$ (Fig.\,\ref{thermalexpansion-b}) reveals two distinct phase transition anomalies at 11.0(1)\,K and 10.6(1)\,K, which coincide with the transition temperatures observed in the various other quantities at $T_{N1}$ and $T_{N2}$, respectively. A closer inspection of the $c$-axis data $\alpha_c$ (cf.\, inset of Fig.\,\ref{thermalexpansion-b} and Fig.\,\ref{thermalexpansion-c}), where both transitions can be separated more easily, discloses a distinctly different character of the two transitions: while the feature at $T_{N1}$ is more step-like, reminiscent of a mean-field transition, the one at $T_{N2}$ has a distinct $\lambda$ shape, indicating substantial contributions from critical fluctuations. The distinction of the two transitions is similar as obsrved in $C_p$ measurements. Upon applying the magnetic field of 6\,T along the $b$-axis, $\alpha_b$ changes significantly in that there is only a single, large $\lambda$-like transition left. Its character seems to suggest that it is a continuation of the transition at $T_{N2}$ - a conjecture which is consistent with the evolution of the features at $T_{N1}$ and $T_{N2}$ seen in magnetic susceptibility and neutron diffraction measurements. Surprisingly, despite the large lattice effects observed in field, there are no accompanying signatures in the dielectric constant.

The results of the magnetic susceptibility for $B||c$ are consistent with thermal expansion data taken along the $c$-axis for the same field orientation, shown in Fig.\,\ref{thermalexpansion-c}. The phase transition anomalies in $\alpha_{c}$ sit on top of a negative background contribution, which is assigned to short-range magnetic correlations (see below). Owing to the pronounced signatures at $T_{N2}$ = 10.6(1)\,K in $\alpha_{c}$, as compared to the small peak in $\alpha_{b}$ (cf.\,Fig.\,\ref{thermalexpansion-b}), the distinct $\lambda$-type character of this transition comes to the fore. The preceding transition at $T_{N1}$ = 11.0(1)\,K, by contrast, features a step-like change, indicative of a more mean-field type transition, cf. discussion above. In a magnetic field of 6\,T, applied parallel to the $ \emph{c}$-axis, both transitions keep their character and shift to higher temperatures by about the same value $\sim$ 0.3\,K, in agreement with the $C_p$ measurements, shown in Fig.\,\ref{fig7}.

\subsection{Short-range ordering effects}

\begin{figure}[!]
\includegraphics[width=86mm,keepaspectratio=true]{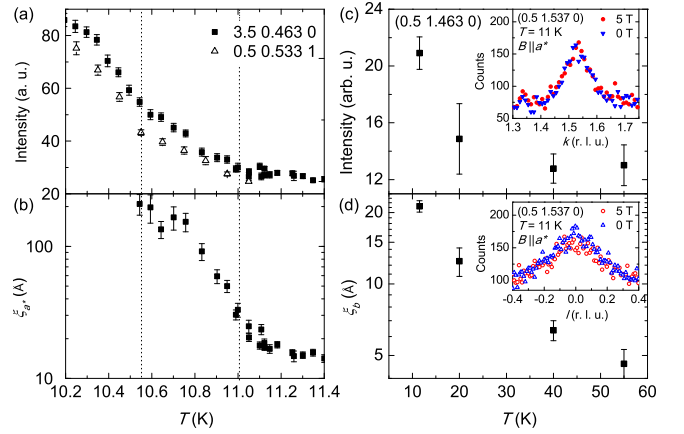}
\caption {(a) Temperature dependences of the (3.5~0.463~0) and the (0.5~0.533~1) magnetic peak intensities measured in the $h$-scan mode and (b) the magnetic correlation length $\xi$ along the $a^*$-axis determined from the magnetic peak width at half maximum, as described in the text. (c) Temperature dependence of the (0.5~1.463~0) magnetic peak intensity measured in the $k$-scan mode and (d) the corresponding magnetic correlation length $\xi$ along $b$. Insets: (c) $k$- and (d) $l$-scans of the (0.5~1.537~0) magnetic peak measured at 11\,K (just above $T_{N1}$) at 0\,T and at 5\,T with $B||a^*$.}
\label{fig8}
\end{figure}

Before discussing the  phase diagram, let us focus first on the temperature interval between $T_{N1}$ and the maximum in the magnetic susceptibility \cite{Becker} ($T_{N} < T < 5T_{N}$), where short-range magnetic correlations are expected to play an important role. The short-range ordering effects have already been observed by $\mu$SR experiments \cite{ZorkoMuSR} and were found to be visible at least up to 20\,K. Our present investigation is based on the neutron diffraction measurements. Within the spherical approximation \cite{Collins, Warren} the magnetic correlation length, $\xi_i$, can be determined from the Lorentzian width of the magnetic reflections as $\xi_i = x_i/(\pi \,\delta_{fwhm_i})$. Here $\delta_{fwhm_i}$ is the magnetic peak width at half maximum given in reciprocal lattice units (r.l.u.) and $x_i$ is the $i$-th unit cell parameter also given in r.l.u. In our experiment, we measured the temperature dependence of selected magnetic reflections (Fig.\,\ref{fig8}). The data were collected in $h$-scan and $k$-scan modes in order to obtain information about the correlations between as well as within the crystal layers. Focusing first on the interlayer correlations, one needs to compare the estimated correlation length along $a^*$, $\xi_{a^*}$, (obtained from an $h$-scan) with the shortest interlayer Fe-Fe distances, $\sim$10\,\AA. \cite{Becker} This would give us an idea about how far above $T_{N1}$ the neighboring layers are correlated. Apparently, magnetic correlations start to shorten already in the HT-ICM phase, where $\xi_{a^*}$ reduces from $\sim$200\,\AA \,at $T_{N2}$ to $\sim$30\,\AA \,at $T_{N1}$  (Fig.\,\ref{fig8}b). $\xi_{a^*}$ is further reduced with increasing temperature and 0.3\,K above $T_{N1}$  $\xi_{a^*} \sim$ 12\,\AA, i.e., almost the same as the interlayer Fe-Fe distances. On further heating, the magnetic peaks become very broad and indistinguishable from the background scattering. Hence, we assume that the magnetic correlations between the crystal layers are suppressed very soon ($<$ 1\,K) above  $T_{N1}$.

This implies that the remaining short-range ordering effects, anticipated from the magnetic susceptibility behavior and observed by $\mu$SR, \cite{ZorkoMuSR} should originate from the correlations within the crystal layers. Such "in-plane" correlations can be estimated from the temperature dependence of the magnetic peak width determined from the $k$- and $l$-scans (insets to Figs.\,\ref{fig8}c and d). In Fig.\,\ref{fig8}d, we show the temperature dependence of the estimated magnetic correlation length along the $b$-axis, $\xi_{b}$, which is just above $T_{N1}$ (at 11\,K) still $\sim$20\,\AA. Comparing this value to the minimal Fe-Fe inter-tetramer distance 4.76\,\AA, we realize that there are still strong correlations among Fe tetramers within the layers. In spite of the obvious reduction of the correlation length above $T_{N1}$, we are able to follow the magnetic peaks up to $\sim$50\,K.

To estimate also the correlations along the $c$-axis, the $l$-scan of the (0.5 1.537 0) magnetic peak at 11\,K was performed (inset to Figs.\,\ref{fig8}d). The obtained width at half maximum is $\sim$0.35 r.l.u., resulting in a calculated correlation length $\xi_{c} \sim$ 13\,\AA. Comparison of this value with  $\xi_{b} \sim$ 20\,\AA \,also determined at 11\,K, implies that even though the closest Fe-Fe distances along $b$ and $c$ are almost identical, i.e., $\sim$4.76\,\AA, the magnetic correlations are stronger along the $b$-axis. Finally we stress that the magnetic field of 5\,T applied along the $a^*$-axis is not strong enough to have an impact on the magnetic correlations, (insets to Figs.\,\ref{fig8}c and d) which is, considering the strong magnetic interactions ($\sim$10\,K), actually anticipated.

\begin{figure}[!]
\includegraphics[width=86mm,keepaspectratio=true]{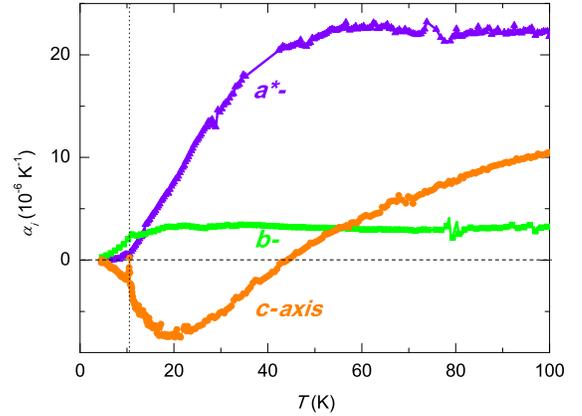}
\caption {Uniaxial thermal expansion coefficients $\alpha_{i}$ measured along the in-plane $b$- and $c$-axis as well as along the out-of plane $a^*$-axis.}
\label{alphai}
\end{figure}

The short-range ordering effects manifest themselves also in the coefficient of thermal expansion. In Fig.\,\ref{alphai} we show the results of all three uniaxial expansion coefficients over an extended temperature range. The data reveal broad anomalies at low temperatures in the in-plane expansion coefficients $\alpha_{c}$ and  $\alpha_{b}$. Upon cooling $\alpha_{c}$ becomes negative below about 50\,K and passes through a broad minimum near 20\,K. At about the same temperature, $\alpha_{b}$ adopts a shallow maximum. Rounded anomalies of this type in the temperature dependence of $\alpha$, which can have either a positive or a negative sign, depending on the pressure dependence of the corresponding characteristic energy, are well-known from short-range magnetic ordering effects. \cite{Bruehl} Since there is no clear corresponding signature in the out-of-plane $\alpha_{a^*}$ data around 20\,K (cf.\, Fig.\,\ref{alphai}), the present results are consistent with the in-plane short-range magnetic ordering setting in around 50\,K.

\section{Discussion and Conclusions}
The prime result of this investigation - a detailed magnetic/electric phase diagram of the FeTe$_2$O$_5$Br multiferroic - is shown in Fig.\,\ref{fig11}. The main features can be summarized as follows:

\begin{figure}[!]
\includegraphics[width=86mm,keepaspectratio=true]{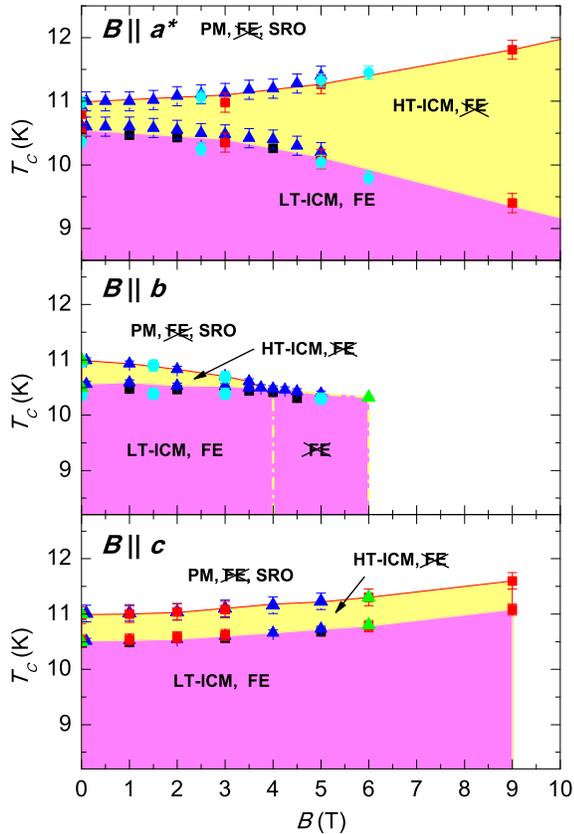}
\caption {Color online: Magnetic phase diagram of the FeTe$_2$O$_5$Br system. Phase transition temperatures as determined from neutron diffraction (blue circles), dielectric constant (black squares), magnetic susceptibility (blue triangles), specific heat (red squares), and thermal expansion (green triangles) measurements.}
\label{fig11}
\end{figure}

1.)~Below $T_{N2}$, the system is in an incommensurate amplitude-modulated phase (LT-ICM), where electric polarization along the $c$-axis is present. This phase was already described in detail in Ref.~\onlinecite{Pregelj}. With increasing temperature, at $T = T_{N2}$ the system undergoes a phase transition into the HT-ICM phase, where electric polarization is lost, while the magnetic order is still incommensurate. With further increasing the temperature, the system undergoes a second phase transition at $T = T_{N1}$, which drives the system into the paramagnetic state. Short-range magnetic correlations within the crystal layers persist up to $\sim$50\,K. By applying an external magnetic field the transition temperatures as well as the phases themselves are significantly altered.

2.)~For $B||a^*$,  $T_{N1}$ increases, whereas $T_{N2}$ decreases with increasing field, implying that $B||a^*$ promotes the HT-ICM phase, while the LT-ICM and the establishment of the electric polarization are being disfavored.

3.)~For $B||b$, $T_{N2}$ seems to be almost unaffected, while $T_{N1}$ slowly decreases, which leads to the extinction of the HT-ICM phase at $B_c \sim$ 4.5\,T. Surprisingly, for fields larger than $B_c$, the electric polarization is also lost - in the entire temperature range, even in the LT-ICM phase. In spite of that, we have not been able to trace any changes in the LT-ICM magnetic ordering.

4.)~For $B||c$, both magnetic transition temperatures shift in parallel towards higher temperatures with increasing magnetic field, indicating that this orientation of the magnetic field is appreciated by the HT-ICM as well as the LT-ICM phases.

Let us now discuss the derived phase diagram and its relevance for the magnetoelectric properties of FeTe$_2$O$_5$Br. We first focus on the HT-ICM phase. In the magnetic field of 6\,T applied along the $a^*$-axis, the temperature interval of the HT-ICM phase is broad enough to firmly  state that the magnetic peak intensity $I$ in the HT-ICM phase has a fundamentally different temperature dependence compared to the LT-ICM phase. Typically, the magnetic peak intensity $I$ is assumed to be proportional to the square of the magnetic moment,\cite{Lawes08} hence its temperature dependence is often described as $I \sim |T - T_{N}|^{2\beta}$. Assuming this simple phenomenological model, we obtain $\beta \sim$ 0.27(1) for the LT-ICM phase and $\beta \sim$ 0.34(2) for the HT-ICM (fits are shown as solid lines in Fig.\,\ref{fig2}a). The magnetic peak intensity in the HT-ICM phase seems to be almost proportional to the temperature. A smaller critical exponent $\beta$ below $T_{N2}$ as compared to that below $T_{N1}$, indicative of enhanced critical fluctuations at $T_{N2}$, is consistent with the behavior found in the thermal expansion and specific heat measurements. A comparison of the obtained $\beta$'s to the critical exponent values known for some typical universality classes \cite{Carlin, Bramwell, Chaikin, Kadanoff} implies that the HT-ICM phase should be characterized as three-dimensional planar (3D XY) or Ising model, while the LT-ICM phase rather corresponds to the two-dimensional planar (2D XY) class. Such a behavior is counterintuitive, as one would expect exactly the opposite ordering sequence. In fact, the temperature dependence of $\xi_{a^*}$ (Fig.\,\ref{fig8}b) indicates that in the HT-ICM phase magnetic correlations between the crystal layers have a finite value, implying a more 2D nature of the phase. At the moment, this puzzle remains to be clarified. We point out though that the obtained $\beta$'s should be taken with care, giving only approximate values, which may deviate from the correct values for the critical exponent. Namely, the exact expression for $I$ depends also on the phase shift and orientation of the magnetic moments, which might exhibit different temperature dependent behavior compared to their magnitudes. Therefore a detailed knowledge about the temperature dependent sliding of the amplitude-modulated waves is required too. Nevertheless, significantly different $\beta$'s in the HT-ICM and LT-ICM phases imply that the two magnetic orders, although being both incommensurate, are intrinsically different. Most likely this is the key to understanding the absence of the ferroelectricity in the HT-ICM phase.

Similar to the magnetic peak intensity, the magnetic peak positions in the LT-ICM phase also exhibits critical behavior, i.e., it can be described with the $|T - T_{N}|^{\gamma}$ law, where $\gamma \sim$0.35(1). On the other hand, the magnetic peak position in the HT-ICM phase is temperature independent and seems to be locked at $k$ = 0.4665(3) (Fig.\,\ref{fig2}). Since the shift of the magnetic peak as well as the electric polarization are both observed only in the LT-ICM phase, we suspect that they are somehow correlated. It is possible that the shift of the magnetic peak indicates the changes of the magnetic structure, which are essential for the development of the electric polarization and are very likely associated with the inversion symmetry breaking at $T_{N2}$.

Next we focus on the field dependence of the observed magnetic transition temperatures, which can be explained by considering the Zeeman energy term in the free energy, $-\chi H^2$. This term implies that in the external magnetic field the state with higher $\chi$ is favored. Thus, in the case of a negative (positive) slope of $\chi(T)$ over the magnetic transition, an increase (decrease) of the transition temperature with increasing magnetic field is anticipated. A close inspection of $\chi(T)$ in the vicinity of the magnetic transitions supports the above argument: when $B||c$, $\chi$ increases with decreasing temperature over both transitions (Fig.\,\ref{fig7}b), which is reflected in the increase of $T_{N1}$ as well as $T_{N2}$ with field; for $B||b$, $\chi$ is decreasing during cooling (inset of Fig.\,\ref{fig4}a), in agreement with the reduction of both transition temperatures; finally, for $B||a^*$, $\chi(T)$ at $T_{N1}$ is almost completely flat while it drops below $T_{N2}$ (Fig.\,\ref{fig1}b), which corroborates  the decrease of $T_{N2}$, but does not say much about the behavior of $T_{N1}$.

Comparison of the explored phase diagram to those found in cycloidal and helical multiferroics with strong magnetoelectric coupling \cite{Nmat07, Fiebig05, ScottNature, Kenzelmann05, Lawes05} reveals that they share a common feature. They all exhibit at least two consecutive magnetic transitions, where only the second one is accompanied with the emergence of electric polarization. This reflects the invariance of the free energy under time reversal, which demands that the lowest magnetoelectric coupling term in multiferroics is trilinear, involving at least two magnetic order parameters. However, in contrast to cycloidal and helical structures, where both magnetic order parameters are typically associated with different components of sublattice magnetizations, in FeTe$_2$O$_5$Br it seems that one of the magnetic order parameters is proportional to the amplitude of the modulation waves, while the second one is related to the phase difference between them. \cite{Pregelj} An additional difference is in the field dependence of the first magnetic transition (at $T_{N1}$), which for $B||a^*$ and $B||c$ shifts to higher temperatures (Fig. \,\ref{fig11}), in contrast to the usual behavior, \cite{Kenzelmann06, Kimura05, Kimura07, Heyer06} where the transition from the paramagnetic to the ICM phase is unfavored by the external magnetic field and thus shifts to lower temperatures.

The apparent differences between FeTe$_2$O$_5$Br from cycloidal or helical systems may be responsible for the intriguing response to $B||b$, when at $\sim$4.5\,T the electric polarization in the LT-ICM disappears in parallel with the loss of the HT-ICM phase, even though magnetic susceptibility and neutron diffraction experiments do not indicate a drastic change of the magnetic structure in the LT-ICM phase. This unusual behavior can be explained by several different scenarios: (i) the magnetic structure of the LT-ICM phase changes above 4.5\,T, but the change is below our sensitivity, (ii) the applied magnetic field narrows the energy gap and allows the low-energy excitations, e.g., phasons, to suppress the long-range ferroelectric ordering, (iii) ferroelectric domains are saturated by the external magnetic field, and consequently diminish the peak in the dielectric constant, (iv) the loss of the HT-ICM phase is accompanied with an induced disorder of the ferroelectric state. Further experiments are clearly needed to clarify this important issue.

Finally, we note that pronounced short-range ordering effects, which indicate low-dimensional magnetic ordering, do not promote the ferroelectricity and raise the multiferroic state as suggested in Ref. \onlinecite{ScottNature}. This is most likely due to the fact that magnetoelectric coupling is conditioned by the loss of inversion symmetry, which is broken only after long-range inversion asymmetric magnetic ordering is established.

To summarize, we have investigated the effect of an applied magnetic field on the magnetoelectric properties of the multiferroic FeTe$_2$O$_5$Br system, i.e., the detailed magnetic phase diagram for magnetic fields applied along all three crystal axes. The first sign of the short-range magnetic correlations within the crystal layers appears already at $\sim$50\,K. At $T_{N1} \sim$ 11.0(1)\,K the system undergoes a magnetic phase transition into the incommensurate HT-ICM phase and 0.5\,K lower, at $T_{N2}$, the system undergoes a second phase transition into the incommensurate amplitude-modulated LT-ICM phase accompanied by the spontaneous electric polarization.\cite{Pregelj} The complex sequence of transitions is similar to many cycloidal and helical structures.\cite{Nmat07, Fiebig05, ScottNature, Kenzelmann05, Lawes05} When a magnetic field is applied, the transition temperatures shift, i.e., for $B||a^*$, $T_{N1}$ increases and $T_{N2}$ decreases, whereas for $B||c$, both magnetic transition temperatures shift in parallel towards higher temperatures. In case of  $B||b$ and $B >$ 4.5\,T, the HT-ICM phase disappears along with the electric polarization otherwise present in the LT-ICM phase. The discovery of the system's ability to turn-off the electric polarization when the external magnetic field of $\sim$4.5\,T is applied along the incommensurate direction is certainly the most prominent discovery in this system.

\acknowledgments
AZ and MJ acknowledge that this work was partially performed and financed as a part of EN$\rightarrow$FIST Centre of Excellence, Dunajska 156, SI-1000 Ljubljana, Slovenia.


\begin{thebibliography}{99}
\bibitem{ScottNature} W. Eerenstein et al., {\em Nature} {\bf 442}, 759-765 (2006).
\bibitem{Kimura} T. Kimura et al., {\em Nature} {\bf 426}, 55-58 (2003).
\bibitem{Nmat07} S. W. Cheong and M. Mostovoy, {\em Nature Mater.} {\bf 6}, 13 (2007).
\bibitem{Kenzelmann05} M. Kenzelmann et al., {\em Phys. Rev. Lett.} {\bf 95}, 087206 (2005);
\bibitem{Lawes05} G. Lawes et al., {\em Phys. Rev. Lett.} {\bf 95}, 087205 (2005).
\bibitem{ES} A. B. Harris et al., {\em Phys. Rev.} {\bf B 73}, 184433 (2006).
\bibitem{ES1} L. C. Chapon et al., {\em Phys. Rev. Lett.}, {\bf 93}, 177402 (2004).
\bibitem{ES2} N. Aliouane et al., {\em Phys. Rev.} {\bf B 73}, 020102(R) (2006).
\bibitem{ES3} I. A. Sergienko et al., {\em Phys. Rev. Lett.}, {\bf 97}, 227204 (2006).
\bibitem{IDM} I.A. Sergienko and E. Dagotto, {\em Phys. Rev.} {\bf B 73}, 094434 (2006).
\bibitem{SC} H. Katsura et al., {\em Phys. Rev. Lett.} {\bf 95}, 057205 (2005).
\bibitem{Johnsson} M. Johnsson et al. {\em Chem. Mater.} {\bf 15} 68 (2003).
\bibitem{Seshadri} R. Seshadri and N. A. Hill {\em Chem. Mater.} {\bf13} 2892 (2001).
\bibitem{Pregelj} M. Pregelj et al., {\em Phys. Rev. Lett.} {\bf 103}, 147202 (2009).
\bibitem{Becker} R. Becker et al., {\em J. Am. Chem. Soc.} {\bf 128}, 15469 (2006).
\bibitem{eps1} Z. Kutnjak et al., {\em Nature} {\bf 441}, 956 (2006).
\bibitem{eps2} Z. Kutnjak and R. Blinc, {\em Phys. Rev.} {\bf B 76}, 104102 (2007).
\bibitem{Pott} R. Pott and R. Schefzyk, {\em J. Phys. E} \textbf{16}, 445  (1983).
\bibitem{Zaharko} O. Zaharko et al., {\em J. Phys.: Conf. Ser.} \textbf{211}, 012002 (2010).
\bibitem{Mariano08} M. de Souza, P. Foury-Leylekian, A. Moradpour, J.-P. Pouget, and M. Lang, {\em Phys. Rev. Lett.} \textbf{101}, 216403 (2008).
\bibitem{ZorkoMuSR} A. Zorko et al.,  {\em J. Appl. Phys.} {\bf 107}, 09D906 (2010).
\bibitem{Collins} M. F. Collins, Magnetic Critical Scattering {\em Oxford University Press}, New York, (1989).
\bibitem{Warren} B. E. Warren, X-ray diffraction {\em Dover Publications}, New York, (1990).
\bibitem{Bruehl}A. Br\"uhl, B. Wolf, V. Pashchenko, M. Anton, C. Gross, W. Assmus, R. Valenti, S. Glocke, A. Kl\"umper, T. Saha-Dasgupta, B. Rahaman, and M. Lang, Phys. Rev. Lett. Phys. Rev. Lett. \textbf{99}, 057204 (2007).
\bibitem{Fiebig05} M. Fiebig, {\em J. Phys. D: Appl. Phys.} {\bf 38}, R123-R152 (2005).
\bibitem{Fiebig05} M. Fiebig, {\em J. Phys. D: Appl. Phys.} {\bf 38}, R123-R152 (2005).
\bibitem{Carlin} R. L. Carlin, Magnetochemistry, {\em Springer-Verlag}, Berlin, p. 41 (1986).
\bibitem{Bramwell}  S. T. Bramwell and P. C. W. Holdsworth, {\em J. Phys. Condens. Matter} {\bf 5}, L53 (1993).
\bibitem{Lawes08}  G. Lawes et al., {\em J. Phys.: Condens. Matter } {\bf 20}, 434205 (2008).
\bibitem{Kenzelmann06} M. Kenzelmann et al., {\em Phys. Rev.} {\bf B 74}, 014429 (2006).
\bibitem{Kimura05}  T. Kimura et al., {\em Phys. Rev.} {\bf B 71}, 224425 (2005).
\bibitem{Kimura07}  H. Kimura et al., {\em J. Phys. Soc. Jpn.} {\bf76}, 074076 (2007).
\bibitem{Heyer06}  O. Heyer et al., {\em J. Phys.: Condens. Matter} {\bf18}, L471 (2006).
\bibitem{Chaikin} P. M. Chaikin and T. C. Lubensky, Principles of condensed matter physics, {\em Cambridge University Press}, Cambridge, (1995).
\bibitem{Kadanoff} L. P. Kadanoff \emph{et al.}, Rev. of Mod. Phys. \textbf{39}, 395 (1967).





\end{thebibliography}
\end{document}